\RequirePackage{fix-cm}
\documentclass[smallextended]{svjour3}       
\smartqed  
\usepackage{graphicx}
\usepackage{amssymb}
\usepackage{comment}
\usepackage{mathtools}
\allowdisplaybreaks
\begin{document}

\title{Optical properties of the solar gravitational lens in the presence of the solar corona}

\titlerunning{Optical properties of the solar gravitational lens}        

\author{Slava G. Turyshev         \and
        Viktor T. Toth 
}


\institute{S. G. Turyshev \at
              Jet Propulsion Laboratory, California Institute of Technology,\\
              4800 Oak Grove Drive, Pasadena, CA 91109-0899, USA
           \and
           V. T. Toth \at
              Ottawa, Ontario K1N 9H5, Canada
}

\date{Received: date / Accepted: date}

\maketitle

\begin{abstract}
We investigate the optical properties of the solar gravitational lens (SGL) in the presence of the solar corona. For this, we consider the combined influence of the static spherically symmetric gravitational field of the Sun---modeled within the first post-Newtonian approximation of the general theory of relativity---and of the solar corona---modeled as a generic, static, spherically symmetric free electron plasma. We study the propagation of monochromatic electromagnetic (EM) waves through the solar system and develop a Mie theory that accounts for the refractive properties of the gravitational field of the Sun and that of the free electron plasma in the extended solar system. We establish a compact, closed-form solution to the boundary value problem and demonstrate that the presence of the solar plasma affects all characteristics of an incident unpolarized light. The affected properties include the direction of the EM wave propagation, its amplitude and its phase, leading to a reduction of the light amplification of the SGL and to a broadening of the corresponding point spread function. The wavelength-dependent plasma effect is critically important at radio frequencies, where it drastically reduces both the  amplification factor of the SGL and also its angular resolution. However, for optical and shorter wavelengths, the plasma's contribution to the EM wave leaves the plasma-free optical properties of the SGL practically unaffected. We discuss the applicability of the SGL for direct high-resolution multipixel imaging and spatially-resolved spectroscopy of exoplanets.
\end{abstract}

\section{Introduction}
\label{sec:intro}

It is well-known that when an electromagnetic (EM) wave propagates through a nonmagnetized free electron plasma, there is a complex interaction between the wave and the medium. Becasue of such interaction, depending on the frequency of the EM wave, the electron plasma frequency and the electron elastic collision frequency, the wave is transmitted, reflected or absorbed by the plasma medium \cite{Ginzburg-book-1964,Landau-Lifshitz:1979}. Detailed understanding of these processes became important with the advent of solar system exploration where EM waves are used for tracking and communicating with deep space probes. The effect of the solar plasma on the propagation of radio waves was explored extensively (e.g., \cite{Allen:1947,Muhleman-etal:1977,Tyler-etal:1977,Muhleman-Anderson:1981}). It is now routinely accounted for in any radio link analysis used for communication and/or navigation and for radio science experiments \cite{Giampieri:1994kj,Bertotti-Giampieri:1998,Verma-etal:2013}.

As far as astronomical observations are concerned, plasma acts as a dispersive medium. Light rays passing through plasma deviate from light-like geodesics in a way that depends on the frequency \cite{Synge-book-1960,Landau-Lifshitz:1988}. This effect plays a significant role in geometric optics models of gravitational microlensing \cite{Clegg:1997ya,Deguchi-Watson:1987}. Refraction of EM waves from a distant background radio source by an interstellar plasma lens with a Gaussian profile of free-electron column density could lead to observable effects \cite{Clegg:1997ya}. Because of their practical importance, the properties of geodesics on a plasma background were investigated extensively. As a result, significant literature on general relativistic ray optics in various media is available (for review, \cite{Perlick-book-2000}).

In the context of the optical properties of the solar gravitational lens (SGL) \cite{vonEshleman:1979,Turyshev:2017}, the effects of the solar corona were investigated using a geometric optics approach \cite{Turyshev-Andersson:2002}. These efforts led to understanding that in the immediate vicinity of the Sun, the propagation of radio waves is significantly affected by the solar plasma. Due to its negative refractive index, the solar corona counteracts gravitational deflection of light by bending the light trajectories outwards and effectively pushing the focal area of the SGL to larger heliocentric distances. The propagation of EM waves at optical frequencies, however, is much less affected by the solar plasma \cite{Turyshev-Toth:2018-plasma}, although the plasma contributes a phase shift that depends on the solar impact parameter of a ray of light.

The present paper offers an overview of our investigation of the optical properties of the SGL using a wave theoretical treatment initiated in \cite{Turyshev:2017,Turyshev-Toth:2017} and presented in detail in \cite{Turyshev-Toth:2018a}. We study light propagation on the background of the solar gravitational monopole and introduce light refraction in the solar corona. We use a generic model for the electron number density in the solar corona, used in \cite{Muhleman-etal:1977,Tyler-etal:1977,Bertotti-Giampieri:1998,Verma-etal:2013} (using the geometric optics approximation) and in \cite{Turyshev-Toth:2018-plasma,Turyshev-Toth:2018-grav-shadow} (using a wave-optical treatment), which extended the results of \cite{Turyshev-Toth:2018} to the case of a free electron plasma distribution. Specifically, we explore the plasma effect on light amplification,  the properties of the point-spread function (PSF), and the resulting angular resolution  of the SGL.

This paper is organized as follows:
In Section~\ref{sec:em-waves-gr+pl} we present Maxwell's field equations for the EM field on the background of the solar gravitational monopole and the solar plasma.
In Section~\ref{sec:debye} we initiate a solution of Maxwell's equations in terms of Debye potentials. The Debye potentials themselves are determined by using the eikonal approximation in Section~\ref{sec:eik}. In Section~\ref{sec:EM-field-outside}, we apply the solution to determine the EM field outside the termination shock. Finally, in Section~\ref{sec:IF-region}, the interference region and image forming therein are discussed. Our results and conclusions are presented in Section~\ref{sec:disc}.

\section{EM waves in a static gravitational field in the presence of plasma}
\label{sec:em-waves-gr+pl}

We consider the propagation of monochromatic light emitted by a distant source beyond the solar system and received by a detector at the focal area of the SGL. This light may be approximated as a plane wave. Its propagation is affected by solar gravity and the plasma of the solar corona. Properties of the plasma are known mostly from spacecraft tracking \cite{Muhleman-etal:1977,Tyler-etal:1977,Muhleman-Anderson:1981,Giampieri:1994kj,Bertotti-Giampieri:1998,Verma-etal:2013}.

The magnetic permeability of the solar corona is negligible. The dielectric permittivity of this plasma is defined as \cite{Landau-Lifshitz:1979}
\begin{equation}
 \epsilon (t,{\vec r})=1- \frac{4\pi n_e (t,{\vec r})e^2}{m_e\omega^2}= 1 - \frac{\omega_{\tt p}^{ 2}}{\omega^2}, \qquad {\rm where} \qquad
 \omega^2_p=\frac{4\pi n_e e^2}{m_e},
 \label{eq:eps}
 \end{equation}
where $e$ is the electron charge, $m_e$ is its mass, while $n_e=n_e(t,{\vec r})$ is the electron number density. The quantity $\omega_{\tt p}$ is known as the electron plasma (or Langmuir) frequency, for which a commonly used model appears in the form
\begin{equation}
\omega_{\tt p}^2=\frac{4\pi e^2}{m_e}
\sum_i \alpha_i\Big(\frac{R_\odot}{r}\Big)^{\beta_i},
\label{eq:n_n-ss}
\end{equation}
with $\alpha_2=3.44\times 10^5~{\rm cm}^{-3}$, $\alpha_6=1.55\times 10^8~{\rm cm}^{-3}$, $\alpha_{16}=2.99\times 10^8~{\rm cm}^{-3}$, $\beta_i=i$ and all other $\alpha_i=0$ \cite{Tyler-etal:1977,Turyshev-Toth:2018-plasma,Turyshev-Toth:2018a}. Inside the opaque Sun, we set $n_e=0$. At distances $r>R_\star$ with $R_\star$ being is the distance to the heliopause, the electron number density is dominated by its approximately constant interstellar value, $n_e=n_0$. In addition to being constant, this value is sufficiently small such that its contribution to refraction can be ignored, as can any variability in the electron number density in the heliopause region.
 
To describe the optical properties of the SGL in the post-Newtonian approximation of the general theory of relativity, we use a static harmonic metric. In spherical coordinates $(r,\theta,\phi)$ the line element is given as \cite{Fock-book:1959,Turyshev-Toth:2013}:
\begin{eqnarray}
ds^2&=&u^{-2}c^2dt^2-u^2\big(dr^2+r^2(d\theta^2+\sin^2\theta d\phi^2)\big),
\label{eq:metric-gen}
\end{eqnarray}
where, to the accuracy sufficient to describe light propagation in the solar system, the quantity $u$ has the form
\begin{eqnarray}
u=1+\frac{r_g}{2r}+{\cal O}(r_g^2),
\label{eq:w-PN}
\end{eqnarray}
with $r_g$ being the Schwarzschild radius of the Sun. This allows us to present the vacuum form of Maxwell's equations for the static, spherically symmetric plasma distribution as
{}
\begin{eqnarray}
{\rm curl}\,{\vec D}&=&-  \mu u^2\frac{1}{c}\frac{\partial {\vec B}}{\partial t}+{\cal O}(r_g^2),
\qquad ~{\rm div}\big(\epsilon u^2\,{\vec D}\big)={\cal O}(r_g^2),
\label{eq:rotE_fl*+*}\\[3pt]
{\rm curl}\,{\vec B}&=& \epsilon u^2\frac{1}{c}\frac{\partial {\vec D}}{\partial t}+{\cal O}(r_g^2),
\qquad \quad \,
{\rm div }\big(\mu u^2\,{\vec B}\big)={\cal O}(r_g^2).
\label{eq:rotH_fl*+*}
\end{eqnarray}

Near the solar surface, $r=R_\odot$, our electron density model (\ref{eq:n_n-ss}) yields $\omega_{\tt p}=\sqrt{4\pi n_ee^2/m_e}\sim 1.20\times 10^9~{\rm s}^{-1}$ corresponding to a frequency of $\nu_p=\omega_{\rm p}/2\pi=191$~MHz. For optical frequencies ($\nu=c/\lambda\sim 300$~THz), its contribution is of ${\cal O}(\omega_{\tt p}^2/\omega^2)\sim 4.08\times10^{-13}$ though for radio frequencies, it can be much higher. At the same time, the contribution of the gravitational monopole to the effective index of refraction is $GM_\odot/2c^2r\lesssim 4.25\times 10^{-6}\, (R_\odot/r).$  Therefore, it is sufficient to carry out the necessary analysis up to terms that are linear with respect to gravity and plasma contributions and neglect higher order terms.

\section{Representation of the EM field in terms of Debye potentials}
\label{sec:debye}

In the static, spherically symmetric case, solving (\ref{eq:rotE_fl*+*})--(\ref{eq:rotH_fl*+*}) is straightforward. Following the derivation in \cite{Turyshev-Toth:2017}, we obtain a system of equations in terms of the electric and magnetic Debye potentials \cite{Born-Wolf:1999}, ${}^e\Pi$ and ${}^m\Pi$, for the components of the monochromatic EM field with wavenumber $k=\omega/c$:
{}
\begin{eqnarray}
{{ D}}_r&=&
\frac{1}{\sqrt{\epsilon}u}\Big\{\frac{\partial^2 }{\partial r^2}
\Big[\frac{r\,{}^e{\hskip -1pt}\Pi}{\sqrt{\epsilon}u}\Big]+\Big(\epsilon\mu\,k^2 u^4-\sqrt{\epsilon}u\big(\frac{1}{\sqrt{\epsilon}u}\big)''\Big)\Big[\frac{r\,{}^e{\hskip -1pt}\Pi}{\sqrt{\epsilon}u}\Big]\Big\},
\label{eq:Dr-em0}\\[3pt]
{{  D}}_\theta&=&
\frac{1}{\epsilon u^2r}\frac{\partial^2 \big(r\,{}^e{\hskip -1pt}\Pi\big)}{\partial r\partial \theta}+\frac{ik}{r\sin\theta}
\frac{\partial\big(r\,{}^m{\hskip -1pt}\Pi\big)}{\partial \phi},
\label{eq:Dt-em0}\\[3pt]
{{  D}}_\phi&=&
\frac{1}{\epsilon u^2r\sin\theta}
\frac{\partial^2 \big(r\,{}^e{\hskip -1pt}\Pi\big)}{\partial r\partial \phi}-\frac{ik}{r}
\frac{\partial\big(r\,{}^m{\hskip -1pt}\Pi\big)}{\partial \theta},
\label{eq:Dp-em0}\\[3pt]
{{  B}}_r&=&
\frac{1}{\sqrt{\mu}u}\Big\{\frac{\partial^2}{\partial r^2}\Big[\frac{r\,{}^m{\hskip -1pt}\Pi}{\sqrt{\mu}u}\Big]+\Big(\epsilon\mu\,k^2 u^4-\sqrt{\mu}u\big(\frac{1}{\sqrt{\mu}u}\big)''\Big)\Big[\frac{r\,{}^m{\hskip -1pt}\Pi}{\sqrt{\mu}u}\Big]\Big\},
\label{eq:Br-em0}\\[3pt]
{{  B}}_\theta&=&
-\frac{ik}{r\sin\theta} \frac{\partial\big(r\,{}^e{\hskip -1pt}\Pi\big)}{\partial \phi}+\frac{1}{\mu u^2r}
\frac{\partial^2 \big(r\,{}^m{\hskip -1pt}\Pi\big)}{\partial r\partial \theta},
\label{eq:Bt-em0}\\[3pt]
{{ B}}_\phi&=&
\frac{ik}{r}\frac{\partial\big(r\,{}^e{\hskip -1pt}\Pi\big)}{\partial \theta}+\frac{1}{\mu u^2r\sin\theta}
\frac{\partial^2 \big(r\,{}^m{\hskip -1pt}\Pi\big)}{\partial r\partial \phi},
\label{eq:Bp-em0}
\end{eqnarray}
where  the electric and magnetic Debye potentials ${}^e{\hskip -1pt}\Pi$ and  ${}^m{\hskip -1pt}\Pi$ satisfy the wave equation
\begin{eqnarray}
\Big(\Delta +k^2\big(1+\frac{2r_g}{r}\big)-V_{\tt p}({ r})\Big)\Big[\frac{\Pi}{u}\Big]={\cal O}(r_g^2, r^{-3}),
\label{eq:Pi-eq*0+*+}
\end{eqnarray}
where the plasma potential $V_{\tt p}$ is given by
\begin{equation}
V_{\tt p}({ r})=\frac{\omega_{\tt p}^{ 2}({ r})}{c^2}=\frac{4\pi e^2}{m_ec^2}
\sum_i \alpha_i \Big(\frac{R_\odot}{r}\Big)^{\beta_i}+{\cal O}\big((kR_\odot)^{-2}\big),
\label{eq:plasma-mod}
\end{equation}
and the quantity $\Pi$ represents either the electric Debye potential, $\,{}^e{\hskip -1pt}\Pi/\sqrt{\epsilon}$, or its magnetic counterpart, $\,{}^m{\hskip -1pt}\Pi/\sqrt{\mu}$.

Typically \cite{Born-Wolf:1999}, in spherical polar coordinates,
Eq.~(\ref{eq:Pi-eq*0+*+}) is solved by separating variables \cite{Turyshev-Toth:2017,Turyshev-Toth:2018-plasma}:
\begin{eqnarray}
{\Pi}=\frac{u}{r}R(r)\Theta(\theta)\Phi(\phi),
\label{eq:Pi*}
\end{eqnarray}
with integration constants and coefficients that are determined by boundary conditions. Direct substitution into (\ref{eq:plasma-mod}) yields the usual solution \cite{Born-Wolf:1999} for $\Phi(\phi)$:
{}
\begin{eqnarray}
\Phi_m(\phi)=e^{\pm im\phi}  \quad\rightarrow \quad \Phi_m(\phi)=a_m\cos (m\phi) +b_m\sin (m\phi),
\label{eq:Ph_m}
\end{eqnarray}
where $\beta=m^2$, $m$ is an integer and $a_m$ and $b_m$ are integration constants. The solution for $\Theta(\theta)$ is well known in the form of spherical harmonics. Single-valued solutions exist with ($l>|m|,$ integer), in the form
{}
\begin{eqnarray}
\Theta_{lm}(\theta)&=&P^{(m)}_l(\cos\theta).
\label{eq:Th_lm}
\end{eqnarray}

The equation for the radial function $R(r)$, in turn, takes the form
{}
\begin{eqnarray}
\frac{d^2 R}{d r^2}+\Big(k^2(1+\frac{2r_g}{r})-\frac{\ell(\ell+1)}{r^2}-V_{\tt p}(r)\Big)R&=&{\cal O}(r^2_g,r_g\frac{\omega_{\tt p}^2}{\omega^2}).
\label{eq:R-bar-k*}
\end{eqnarray}

To solve (\ref{eq:R-bar-k*}), we follow \cite{Turyshev-Toth:2018-plasma} and first separate the terms in plasma potential $V_{\tt p}$, (\ref{eq:plasma-mod}), by isolating the $1/r^2$ term and representing the remaining terms as the short-range potential $V_{\tt sr}$:
{}
\begin{eqnarray}
V_{\tt p}(r)&=& \frac{\mu^2}{r^2}+ V_{\tt sr},
\label{eq:V-sr-m2]}
\end{eqnarray}
where $\mu^2$ and $V_{\tt sr}$ are given by (note the reuse of the symbol $\mu$; not to be confused with magnetic permeability, which we no longer use):
{}
\begin{eqnarray}
\mu^2&=&\frac{4\pi e^2R^2_\odot}{m_ec^2} \alpha_2, \qquad {} V_{\tt sr}\,=\,\frac{4\pi e^2}{m_ec^2}
\sum_{i>2} \alpha_i \Big(\frac{R_\odot}{r}\Big)^{\beta_i}+{\cal O}\big((kR_\odot)^{-2}\big).
\label{eq:V-sr-m2}
\end{eqnarray}
From the phenomenological model (\ref{eq:n_n-ss}), we obtain $\mu^2\simeq 5.89\times 10^{15}$.  The range of $V_{\tt sr}$ is very short: this potential provides a negligible contribution after $r\simeq 8 R_\odot$. These terms allow us to present the radial equation (\ref{eq:R-bar-k*}) as
{}
\begin{eqnarray}
\frac{d^2 R_L}{d r^2}+\Big(k^2(1+\frac{2r_g}{r})-\frac{L(L+1)}{r^2}- V_{\tt sr}(r)\Big)R_L&=&{\cal O}\big(r^2_g,r_g\frac{\omega_{\tt p}^2}{\omega^2}\big),
\label{eq:R-bar-k*2}
\end{eqnarray}
where the new index $L$ for the plasma-modified centrifugal potential is determined from
\begin{equation}
L(L+1)=\ell(\ell+1)+\mu^2.
\label{eq:L}
\end{equation}
When $\mu/\ell\ll1$, this solution behaves as
{}
\begin{equation}
L=\ell+\frac{\mu^2}{2\ell+1} +{\cal O}(\mu^{4}/\ell^3).
\label{eq:L2-apr}
\end{equation}
The value of $\ell$ may be estimated using its relation to the classical impact parameter, namely $\ell=kb\geq kR_\odot=4.37\times 10^{15}$. Therefore, we see that the ratio $\mu/\ell\leq 1.75\times 10^{-8}$ is indeed small, justifying the approximation (\ref{eq:L2-apr}).

\section{Eikonal solution for Debye potential}
\label{sec:eik}

No analytical solution is known to exist for Eq.~(\ref{eq:Pi-eq*0+*+}) in the general case when $V_{\tt sr}\ne 0$. Therefore, we seek a suitable approximation method. The eikonal approximation is valid when the following two criteria are satisfied \cite{Sharma-Sommerford-book:2006}: $kb\gg 1$ and $V_{\tt sr}(r)/k^2\ll 1$, where $k$ is the wave number and $b$ is the impact parameter. In our case, the first condition yields $kb=4.37\times 10^{15}\,(\lambda/1\,\mu{\rm m})(b/R_\odot) \gg 1$. Taking the short-range plasma potential $V_{\tt sr}$ from (\ref{eq:V-sr-m2}), we evaluate the second condition as $V_{\tt sr}(r)/k^2\leq V_{\tt sr}(R_\odot)/k^2\approx 4.07\times 10^{-13} \,(\lambda/1\,\mu{\rm m})^2\ll1$. Therefore, we may proceed. We first note that when the short-range potential $V_{\tt sr}$ is absent, (\ref{eq:R-bar-k*2})  takes the form
{}
\begin{eqnarray}
\frac{d^2 R_L}{d r^2}+\Big(k^2(1+\frac{2r_g}{r})-\frac{L(L+1)}{r^2}\Big)R_L&=&0.
\label{eq:R-bar-k*20}
\end{eqnarray}
The well-known solution to this equation is given in terms of the Coulomb functions $F_L(kr_g,kr)$ and $G_L(kr_g,kr)$ \cite{Abramovitz-Stegun:1965,Schiff:1968,Landau-Lifshitz:1989,Messiah:1968,Turyshev-Toth:2017}:
{}
\begin{eqnarray}
R^{(2)}_L=c_LF_L(kr_g,kr)+d_LG_L(kr_g,kr),
\end{eqnarray}
where we use the superscript ${(2)}$ to indicate that the solution to (\ref{eq:R-bar-k*20}) includes the term $\propto 1/r^2$. Combining the solution for $R^{(2)}_L$ with the results for $\Phi(\phi)$, $\Theta(\theta)$, we obtain the corresponding Debye potential:
{}
\begin{align}
\hskip -0.75em
\Pi^{(2)}({\vec r})=\frac{1}{r} \sum_{\ell=0}^\infty\sum_{m=-\ell}^\ell
\mu_\ell R^{(2)}_L(r)\big[ P^{(m)}_l(\cos\theta)\big]\big[a_m\cos (m\phi) +b_m\sin (m\phi)\big],
\label{eq:Pi-degn-sol-00}
\end{align}
with the constants $\mu_\ell$, $a_m$ and $b_m$  to be determined later. We see that $\Pi^{(2)}({\vec r})$ is a solution to the following wave equation:
{}
\begin{eqnarray}
\Big(\Delta +k^2(1+\frac{2r_g}{r})-\frac{\mu^2}{r^2}\Big)\Pi^{(2)}({\vec r})=0,
\label{eq:Pi-eq*0+*+0}
\end{eqnarray}
which is the equation for the ``free'' Debye potential in the presence of gravity and $1/r^2$ plasma, $\Pi^{(2)}({\vec r})$, and which is yet ``unperturbed'' by the short-range plasma potential, $V_{\tt sr}$.

To solve (\ref{eq:Pi-eq*0+*+}), we first write this equation as
{}
\begin{eqnarray}
\Big(\Delta +k^2(1+\frac{2r_g}{r})-\frac{\mu^2}{r^2}-V_{\tt sr}({ r})\Big)\Pi({\vec r})={\cal O}(r_g^2, r^{-3}).
\label{eq:Pi-eq*0+*+1*}
\end{eqnarray}
We consider a trial solution in the form
{}
\begin{eqnarray}
\Pi({\vec r})=\Pi^{(2)}({\vec r})\phi(\vec r),
\label{eq:Pi-eq*0+*+1}
\end{eqnarray}
In other words, the Debye potential $\Pi^{(2)}(\vec r)$, becomes ``distorted'' in the presence of the potential $V_{\tt sr}$ by $\phi$, a slowly varying function of $r$. This enables us to use the eikonal approximation, to arrive (after some algebra) at the solution
\begin{equation}
\Pi(\vec r)=\Pi^{(2)}(\vec r)\exp\Big\{\pm i \xi_b(\tau) \Big\}+{\cal O}(\omega^4_{\tt p}/\omega^4),
\label{eq:eik6h}
\end{equation}
with the eikonal phase given by
{}
\begin{eqnarray}
\xi_b(r)
&=& -\frac{2\pi e^2R_\odot}{m_ec^2k}
\sum_{i>2}\alpha_i\Big(\frac{R_\odot}{b}\Big)^{\beta_i-1}\Big\{Q_{\beta_i}(\tau)-Q_{\beta_i}(\tau_0)\Big\},
\label{eq:delta-D*-av0WKB+1*}
\end{eqnarray}
where we introduced the function $Q_{\beta_i}(\tau)$, which, with $\tau=({\vec k}\cdot {\vec r})=\sqrt{r^2-b^2}$, is given as
\begin{eqnarray}
Q_{\beta_i}(\tau)={}_2F_1\Big[{\textstyle\frac{1}{2}},{\textstyle\frac{1}{2}}\beta_i,{\textstyle\frac{3}{2}},-\frac{\tau^2}{b^2}\Big]\frac{\tau}{b},
\label{eq:Q}
\end{eqnarray}
with ${}_iF_j[...,z]$ being the hypergeometric function \cite{Abramovitz-Stegun:1965}. For $r=b$ or, equivalently, for $\tau=0$, the function (\ref{eq:Q}) is well-defined, taking the value of $Q_{\beta_i}(0)=0$, for each $\beta_i$. For large values of $r$ and, thus for large $\tau$, for any given value of $\beta_i$, the function $Q_{\beta_i}(\tau)$ rapidly approaches a limit:
\begin{equation}
\lim\limits_{\tau\to\infty}Q_{\beta_i}\big(\tau\big)
=\frac{{\textstyle\frac{1}{2}}\beta_i}{\beta_i-1}B[{\textstyle\frac{1}{2}}\beta_i+{\textstyle\frac{1}{2}},{\textstyle\frac{1}{2}}],
\label{eq:eik1hQ}
\end{equation}
with $B[x,y]$ being Euler's beta function \cite{Turyshev-Toth:2018-plasma}. For the values of $\beta_i$ used in the model (\ref{eq:n_n-ss}) for the electron number density in the solar corona, $\beta_i=\{2,6,16\}$, these values are $Q^\star_{2,6,16}=\{\frac{1}{2}\pi,\frac{3}{16}\pi,\frac{429}{4096}\pi\}$, respectively. The quantities $Q_{\beta_i}(r)$ (\ref{eq:Q}) for $\beta_i>2$ are always small, $0\leq |Q_{\beta_i}|<1$, and as functions of $r$, they reach their asymptotic values $Q^\star_{\beta_i}$ (\ref{eq:eik1hQ}) quite rapidly, typically after $r\simeq3.2 b$.

We may thus present $R_L$ in the following form:
{}
\begin{eqnarray}
R_L (r) &=&  \cos\xi_b(r)F_L(kr_g, kr) + \sin\xi_b(r)\,G_L(kr_g, kr),
\label{eq:R-L3}
\end{eqnarray}
which explicitly shows the phase shift, $\xi_b(r)$, induced by the short-range plasma potential, satisfying relevant boundary conditions \cite{Burke-book-2011}. This expression, together with (\ref{eq:Pi*}), determines the electromagnetic field inside the termination shock boundary, where the plasma is characterized by the model (\ref{eq:n_n-ss}).

\section{General solution for the EM field outside the termination shock}
\label{sec:EM-field-outside}

Outside the termination shock, $r>R_\star$, we model the solution for the Debye potential, as usual, as a combination of that of a Coulomb-modified incident plane wave (with the $G_L$ term omitted since it becomes infinite at the origin) and a scattered wave. These two solutions must be consistent on the boundary, that is, at $r=R_\star$. On the other hand, the scattered wave must vanish at infinity. The Coulomb--Hankel functions $H^+_L(kr_g,kr)=G_L(kr_g,kr)+iF_L(kr_g,kr)$ impart precisely this property. After a significant amount of algebra, we arrive at the solution for the Debye potential outside the termination shock:
\begin{align}
\label{eq:Pi-s_a1*0}
\Pi_{\tt out}(r, \theta)=\frac{E_0}{k^2}\frac{u}{r}&\sum_{\ell=1}^\infty i^{\ell-1}\frac{2\ell+1}{\ell(\ell+1)}e^{i\sigma_\ell}\\
&{}\times\Big\{F_\ell(kr_g,kr) + \frac{1}{2i}\Big(e^{2i\delta_\ell^\star}-1\Big) H^+_\ell(kr_g, kr)\Big\}P^{(1)}_\ell(\cos\theta),\nonumber
\end{align}
while the matching potential inside the termination shock is given by
\begin{align}
\label{eq:Pi-in+sl}
\Pi_{\tt in}(r, \theta)=\frac{E_0}{k^2} \frac{u}{r}&\sum_{\ell=1}^\infty i^{\ell-1}\frac{2\ell+1}{\ell(\ell+1)}e^{i\big(\sigma_\ell +\delta_\ell^\star -\delta_\ell(r)\big)}\\
&{}\times\Big\{F_\ell(kr_g, kr)+ \frac{1}{2i}\Big(e^{2i\delta_\ell(r)}-1\Big)H^+_\ell(kr_g, kr)\Big\}P^{(1)}_\ell(\cos\theta),\nonumber
\end{align}
with all constants determined by the boundary conditions. In these expressions, the plasma-induced phaseshift is given by
\begin{align}
\delta_\ell(r)&=-\frac{\pi}{2}(L-\ell)+\sigma_L-\sigma_\ell+\xi_b(r),
\end{align}
$\delta_\ell^\star=\delta_\ell(R^\star)$, and where the Coulomb-type phaseshift due to gravitation is
\begin{align}
\sigma_\ell&=\arg \Gamma(\ell+1-ikr_g)=\sigma_0-\sum_{j=1}^\ell\arctan\dfrac{kr_g}{j},
\label{eq:S-l-g-s}
\end{align}
where for $kr_g\rightarrow\infty$ the constant $\sigma_0=\arg\Gamma(1-ikr_g)$ is given by the expression $\sigma_0= -kr_g\ln {kr_g}/{e}-{\pi}/{4}$ \cite{Turyshev-Toth:2017}. Replacing the sum in (\ref{eq:S-l-g-s}) with an integral, for $\ell\gg kr_g$, we obtain $\sigma_\ell$  \cite{Turyshev-Toth:2018-grav-shadow}:
{}
\begin{eqnarray}
\sigma_\ell&=& -kr_g\ln \ell.
\label{eq:sig-l*}
\end{eqnarray}

Thus, we have identified all the Debye potentials involved in the Mie problem. However, the presence of the Sun itself is not yet captured. For this, similarly to \cite{Turyshev-Toth:2017,Turyshev-Toth:2018-plasma}, we apply the fully absorbing boundary conditions that represent the physical size and the surface properties of the Sun \cite{Turyshev-Toth:2018-grav-shadow}.To set the boundary conditions, we rely on the semiclassical analogy between the partial momentum, $\ell$, and the impact parameter, $b$, that is given as $\ell=kb$ \cite{Messiah:1968,Landau-Lifshitz:1989}. We require that rays with impact parameters $b\le R_\odot^\star=R_\odot +r_g$ are completely absorbed by the Sun \cite{Turyshev-Toth:2017}.

Relying on the representation of the regular Coulomb function $F_\ell$ via incoming, $H^{+}_\ell$, and outgoing, $H^{-}_\ell$, waves as $F_\ell=(H^{+}_\ell-H^{-}_\ell)/2i$ (discussed in \cite{Turyshev-Toth:2017}), we may express the Debye potential (\ref{eq:Pi-s_a1*0}) as
{}
\begin{align}
\Pi(r, \theta)= \frac{E_0}{2ik^2}\frac{u}{r}&\sum_{\ell=1}^\infty i^{\ell-1}\frac{2\ell+1}{\ell(\ell+1)}e^{i\sigma_\ell}\nonumber\\
&{}\times\Big\{e^{2i\delta_\ell^\star}H^+_\ell(kr_g, kr)-H^-_\ell(kr_g, kr)\Big\}P^{(1)}_\ell(\cos\theta).
\label{eq:Pi-s_a*0}
\end{align}

This form of the combined Debye potential is convenient for implementing the fully absorbing  boundary condition. Subtracting from (\ref{eq:Pi-s_a*0})  the outgoing wave (i.e., $\propto H^{(+)}_\ell$) for the impact parameters $b\leq R_\odot^\star$ or equivalently for $\ell\in[1,kR_\odot^\star]$, we obtain
 {}
\begin{align}
\Pi(r, \theta)=&\frac{E_0}{2ik^2}\frac{u}{r}\sum_{\ell=1}^\infty i^{\ell-1}\frac{2\ell+1}{\ell(\ell+1)}e^{i\sigma_\ell}\nonumber\\
&\hskip 3em{}\times\Big\{e^{2i\delta_\ell}H^+_\ell(kr_g, kr)-H^-_\ell(kr_g, kr)\Big\}P^{(1)}_\ell(\cos\theta)\nonumber\\
&{}-\frac{E_0}{2ik^2}\frac{u}{r}\sum_{\ell=1}^{kR_\odot^\star} i^{\ell-1}\frac{2\ell+1}{\ell(\ell+1)}e^{i\sigma_\ell}e^{2i\delta_\ell^\star}H^+_\ell(kr_g, kr)P^{(1)}_\ell(\cos\theta).
\label{eq:Pi-s_a+0}
\end{align}

This is our main result, valid for all distances outside the termination shock $r>R_\star$ and all angles. It requires the tools of numerical analysis to fully explore its behavior and the resulting EM field \cite{Kerker-book:1969,vandeHulst-book-1981,Grandy-book-2005}. However, for our purposes, we need to know the field in the forward direction. Furthermore, our main interest is to study the largest plasma impact on light propagation, which corresponds to the smallest values of the impact parameter. We thus may simplify the result (\ref{eq:Pi-s_a+0}) by taking into account the asymptotic behavior of the function $H^{+}_\ell(kr_g,kr)$, considering the field at large heliocentric distances, $kr\gg\ell$ (see p.~631 of \cite{Morse-Feshbach:1953}). Such an expression is given in the form
{}
\begin{align}
\lim_{kr\rightarrow\infty} H^{+}_\ell(kr_g,kr)\sim
\exp\Big[&i\Big(k(r+r_g\ln 2kr)+\frac{\ell(\ell+1)}{2kr}
+\frac{[\ell(\ell+1)]^2}{24k^3r^3}\nonumber\\
&{}+\sigma_\ell-\frac{\pi\ell}{2}\Big)\Big] +{\cal O}\big((kr)^{-5}, r_g^2\big),
\label{eq:Fass*}
\end{align}
which includes the contribution from the centrifugal potential in the radial equation (\ref{eq:R-bar-k*}) (see also Appendix A in \cite{Turyshev-Toth:2018} or \cite{Kerker-book:1969}).

Using (\ref{eq:Fass*}), we can present the solution for the Debye potential outside the termination shock, for $r>R_\star$ to ${\cal O}(r^2_g,r_g{\omega_{\tt p}^2}/{\omega^2})$  in the following form
{}
\begin{align}
\Pi (r, \theta)=&\Pi_0 (r, \theta)+\frac{ue^{ik(r+r_g\ln 2kr)}}{r}\frac{E_0}{2k^2}\nonumber\\
&\hskip -1.5em{}\times \Big\{\sum_{\ell=1}^{kR_\odot^\star} \frac{2\ell+1}{\ell(\ell+1)}e^{i\big(2\sigma_\ell+\frac{\ell(\ell+1)}{2kr}+\frac{[\ell(\ell+1)]^2}{24k^3r^3}\big)}P^{(1)}_\ell(\cos\theta)-{}\nonumber\\
&\hskip -1em {}- \sum_{\ell=kR_\odot^\star}^{\infty} \frac{2\ell+1}{\ell(\ell+1)}e^{i\big(2\sigma_\ell+\frac{\ell(\ell+1)}{2kr}+\frac{[\ell(\ell+1)]^2}{24k^3r^3}\big)}\big(e^{i2\delta_\ell^\star}-1\big)P^{(1)}_\ell(\cos\theta) \Big\}.
\label{eq:Pi_g+p0}
\end{align}

The first term in (\ref{eq:Pi_g+p0}), $\Pi_0 (r, \theta)$, is the Debye potential representing the incident EM wave propagating in the vacuum on the background of a post-Newtonian field of a gravitational monopole. It is convenient to use an exact expression for $\Pi_0$, which was derived in \cite{Turyshev-Toth:2017} in the form
{}
\begin{align}
\label{eq:sol-Pi0*}
\Pi_0(r, \theta)=&-\psi_0\frac{iu}{k}\frac{1-\cos\theta}{\sin\theta}\\
&{}\times\Big(e^{ikz}{}_1F_1[1+ikr_g,2,ikr(1-\cos\theta)]-e^{-ikr}{}_1F_1[1+ikr_g,2,2ikr]\Big),\nonumber
\end{align}
where $\psi^2_0= E_0^2\,{2\pi kr_g}/({1-e^{-2\pi kr_g}})$.

With the solution for the Debye potential given by (\ref{eq:Pi_g+p0}), and with the help of (\ref{eq:Dr-em0})--(\ref{eq:Bp-em0}) (also see \cite{Turyshev-Toth:2017}), we may now compute the EM field in the various regions involved. Given the smallness of the ratio $(\omega_{\tt p}/\omega)^2$ ($\sim 10^{-2}$ for radio and $\sim 10^{-11}$ for optical wavelengths), and especially at large heliocentric distances, we may neglect the distance-dependent effect of the solar plasma on the amplitude of the EM wave. Therefore, we can put $\epsilon =\mu =1$ in (\ref{eq:Dr-em0})--(\ref{eq:Bp-em0}) and use the following expressions to construct the EM field in the static, spherically symmetric geometry (see details in \cite{Turyshev-Toth:2017}):
{}
\begin{align}
  \left( \begin{aligned}
{   D}_r& \\
{   B}_r& \\
  \end{aligned} \right) =&  \left( \begin{aligned}
\cos\phi \\
\sin\phi  \\
  \end{aligned} \right) \,e^{-i\omega t}\alpha(r, \theta), &
    \left( \begin{aligned}
{   D}_\theta& \\
{   B}_\theta& \\
  \end{aligned} \right) =&  \left( \begin{aligned}
\cos\phi \\
\sin\phi  \\
  \end{aligned} \right) \,e^{-i\omega t}\beta(r, \theta),
\nonumber\\
    \left( \begin{aligned}
{   D}_\phi& \\
{   B}_\phi& \\
  \end{aligned} \right) =&  \left( \begin{aligned}
-\sin\phi \\
\cos\phi  \\
  \end{aligned} \right) \,e^{-i\omega t}\gamma(r, \theta),
  \label{eq:DB-sol00p*}
\end{align}
with $\alpha, \beta$ and $\gamma$ computed from the known Debye potential, $\Pi$, as
{}
\begin{eqnarray}
\alpha(r, \theta)&=&
\frac{1}{u}\Big\{\frac{\partial^2 }{\partial r^2}
\Big[\frac{r\,{\hskip -1pt}\Pi}{u}\Big]+k^2 u^4\Big[\frac{r\,{\hskip -1pt}\Pi}{u}\Big]\Big\}+{\cal O}\Big(\big(\frac{1}{u}\big)''\Big),
\label{eq:alpha*}\\
\beta(r, \theta)&=&\frac{1}{u^2r}
\frac{\partial^2 \big(r\,{\hskip -1pt}\Pi\big)}{\partial r\partial \theta}+\frac{ik\big(r\,{\hskip -1pt}\Pi\big)}{r\sin\theta},
\label{eq:beta*}\\[0pt]
\gamma(r, \theta)&=&\frac{1}{u^2r\sin\theta}
\frac{\partial \big(r\,{\hskip -1pt}\Pi\big)}{\partial r}+\frac{ik}{r}
\frac{\partial\big(r\,{\hskip -1pt}\Pi\big)}{\partial \theta}.
\label{eq:gamma*}
\end{eqnarray}

\begin{figure}[t]
\includegraphics[width=\linewidth]{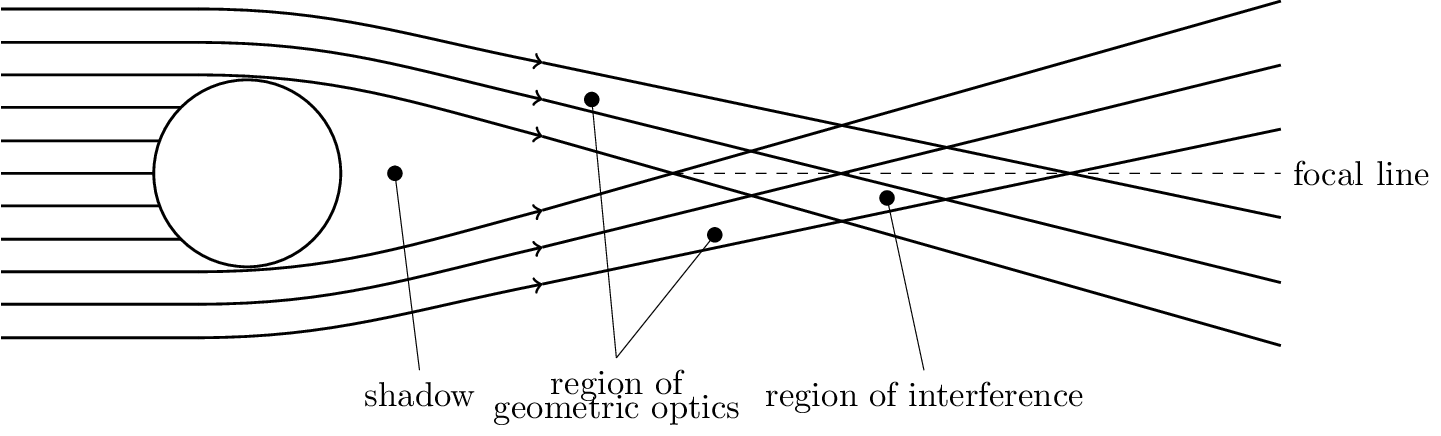}
\caption{
Three different regions of space associated with a monopole gravitational lens: the shadow, the region of geometric optics, and the region of interference (from \cite{Turyshev-Toth:2017}).\label{fig:regions}}
\end{figure}

In the shadow (Fig.~\ref{fig:regions}) behind the Sun (i.e, for impact parameters $b\leq R^\star_\odot$) the EM field is represented by the Debye potential of the shadow, $\Pi_{\tt sh}$, given as
{}
\begin{align}
\label{eq:Pi_sh}
\Pi_{\tt sh} (r, \theta)=&
\Pi_0 (r, \theta)+\frac{ue^{ik(r+r_g\ln 2kr)}}{r}\frac{E_0}{2k^2}\\
&{}\times\sum_{\ell=1}^{kR_\odot^\star} \frac{2\ell+1}{\ell(\ell+1)}e^{i\big(2\sigma_\ell+\frac{\ell(\ell+1)}{2kr}+\frac{[\ell(\ell+1)]^2}{24k^3r^3}\big)}P^{(1)}_\ell(\cos\theta)+{\cal O}(r^2_g,r_g\frac{\omega_{\tt p}^2}{\omega^2}).\nonumber
\end{align}
As was shown in \cite{Turyshev-Toth:2018-grav-shadow}, the potential $\Pi_{\tt sh} (r, \theta)$ yields no EM filed in this region.

In the region behind the Sun but outside the solar shadow (i.e., for light rays with impact parameters $b>R_\odot$), which includes both the geometric optics and interference regions (in the immediate vicinity of the focal line), the EM field is derived from the Debye potential given by the remaining terms in (\ref{eq:Pi_g+p0}) to ${\cal O}(r^2_g,r_g{\omega_{\tt p}^2}/{\omega^2})$ as
{}
\begin{eqnarray}
{\Pi (r, \theta)}&=&
{\Pi_0 (r, \theta)}+{\Pi_{\tt p} (r, \theta)},
\label{eq:Pi-g+p}
\end{eqnarray}
where the plasma scattering Debye potential  ${\Pi_{\tt p} (r, \theta)}$ is given by
{}
\begin{align}
\Pi_{\tt p}(r,\theta)=&-\frac{ue^{ik(r+r_g\ln 2kr)}}{r}\frac{E_0}{2k^2}\sum_{\ell=kR_\odot^\star}^\infty \frac{2\ell+1}{\ell(\ell+1)}e^{i\big(2\sigma_\ell+\frac{\ell(\ell+1)}{2kr}+\frac{[\ell(\ell+1)]^2}{24k^3r^3}\big)}\nonumber\\
&\hskip 3em{}\times \Big(e^{i2\delta_\ell^\star}-1\Big)
P^{(1)}_\ell(\cos\theta).
\label{eq:f-v*+}
\end{align}

Expression (\ref{eq:Pi-g+p}) with (\ref{eq:f-v*+}) is our main result for the regions outside the termination shock, $r>R_\star$ and also outside the shadow region, i.e., $b\geq R_\odot^\star$. We now substitute (\ref{eq:f-v*+})  in the expressions (\ref{eq:alpha*})--(\ref{eq:gamma*}) to derive the factors $\alpha(r,\theta), \beta(r,\theta)$ and $\gamma(\theta)$ characterizing the EM field produced by the plasma potential $\Pi_{\tt p}(r,\theta)$, which to ${\cal O}\big(r^2_g,r_g({\omega_{\tt p}^2}/{\omega^2})\big)$ are given as
{}
\begin{align}
\label{eq:alpha*1*}
\alpha(r,\theta) =& -E_0\frac{e^{ik(r+r_g\ln 2kr)}}{uk^2r^2}\sum_{\ell=kR^\star_\odot}^\infty(\ell+{\textstyle\frac{1}{2}})e^{i\big(2\sigma_\ell+\frac{\ell(\ell+1)}{2kr}+\frac{[\ell(\ell+1)]^2}{24k^3r^3}\big)}\\
&\hskip 3em{}\times\Big(e^{i2\delta^*_\ell}-1\Big)P^{(1)}_\ell(\cos\theta){\cal A}_\ell(r),
\nonumber\\
\label{eq:beta*1*}
\beta(r,\theta) =& E_0\frac{ue^{ik(r+r_g\ln 2kr)}}{ikr}\sum_{\ell=kR^\star_\odot}^\infty\frac{\ell+{\textstyle\frac{1}{2}}}{\ell(\ell+1)}e^{i\big(2\sigma_\ell+\frac{\ell(\ell+1)}{2kr}+\frac{[\ell(\ell+1)]^2}{24k^3r^3}\big)}\\
&\hskip 3em{}\times\Big(e^{i2\delta^*_\ell}-1\Big)\Big\{\frac{\partial P^{(1)}_\ell(\cos\theta)}{\partial \theta}{\cal B}_\ell(r)
+\frac{P^{(1)}_\ell(\cos\theta)}{\sin\theta}\Big\},\nonumber\\
\label{eq:gamma*1*}
\gamma(r,\theta) =& E_0\frac{ue^{ik(r+r_g\ln 2kr)}}{ikr}\sum_{\ell=kR^\star_\odot}^\infty\frac{\ell+{\textstyle\frac{1}{2}}}{\ell(\ell+1)}e^{i\big(2\sigma_\ell+\frac{\ell(\ell+1)}{2kr}+\frac{[\ell(\ell+1)]^2}{24k^3r^3}\big)}\\
&\hskip 3em{}\times\Big(e^{i2\delta^*_\ell}-1\Big)\Big\{\frac{\partial P^{(1)}_\ell(\cos\theta)}{\partial \theta}+\frac{P^{(1)}_\ell(\cos\theta)}{\sin\theta}{\cal B}_\ell(r)
\Big\}.\nonumber
\end{align}
where ${\cal A}_\ell(r)$ and ${\cal B}_\ell(r)$ are given by
\begin{eqnarray}
{\cal A}_\ell(r)&=&u^{2}+\frac{\ell(\ell+1)}{4k^2r^2}\frac{r_g}{r}+\frac{i}{kr}\Big(1+\frac{\ell(\ell+1)}{2k^2r^2}\Big)-\frac{ikr_g}{\ell(\ell+1)}+
{\cal O}\big((kr)^{-4}\big),~~~~\\
{\cal B}_\ell(r)&=&1-u^{-2}\Big(\frac{\ell(\ell+1)}{2k^2r^2}+\frac{[\ell(\ell+1)]^2}{8k^4r^4}\Big)+\frac{ir_g}{2kr^2}+{\cal O}\big((kr)^{-5}\big).
\label{eq:A*B*}
\end{eqnarray}

This is an important result that allows us to describe the EM field in all the regions of interest for the SGL.
In particular, in accord with (\ref{eq:Pi-g+p}), the total EM field in the region of geometrical optics is given by the sum of the incident and scattered EM waves. Up to terms of ${\cal O}(\theta^5,\delta\theta^2_{\tt p},r_g\theta^4/r,r_g^2,{r\delta\theta_{\tt p}}/{b})$ the components of this field have the form:
{}
\begin{align}
  \left( \begin{aligned}
{    D}_r& \\
{    B}_r& \\
  \end{aligned} \right) =&
  E_0u^{-1}
\sin\Big(\theta\pm\frac{d2\delta^\star_\ell}{d\ell}\Big)\Big(1+\frac{r_g}{r(1-\cos\theta)}\Big) \left( \begin{aligned}
\cos\phi \\
\sin\phi  \\
  \end{aligned} \right)\,e^{i\psi_{\tt go}(r,\theta,t)},
  \label{eq:DB-t-pl=pV0}\\
  \left( \begin{aligned}
{    D}_\theta& \\
{    B}_\theta& \\
  \end{aligned} \right) =&
  E_0u^{-1}
  \Big(\cos\big(\theta\pm\frac{d2\delta^\star_\ell}{d\ell}\big)-\frac{r_g}{r}\Big)\left( \begin{aligned}
\cos\phi \\
\sin\phi  \\
  \end{aligned} \right) \,e^{i\psi_{\tt go}(r,\theta,t)},
  \label{eq:DB-t-pl=pV1}\\
\left( \begin{aligned}
{    D}^{\tt }_\phi& \\
{    B}^{\tt }_\phi& \\
  \end{aligned} \right) =&
  E_0u
\left( \begin{aligned}
-\sin\phi \\
\cos\phi  \\
  \end{aligned} \right) \,e^{i\psi_{\tt go}(r,\theta,t)}.
  \label{eq:DB-t-pl=pV2}
\end{align}
with  $\psi_{\tt go}$ is the phase in the geometric optics region that is given as follows $\psi_{\tt go}(r,\theta,t)=kr\cos\theta-kr_g\ln kr(1-\cos\theta)+2\delta^*_\ell-\omega t$. Also, the plasma phase shift, $\delta_\ell^*$, for the model (\ref{eq:n_n-ss}) is given as
{}
\begin{eqnarray}
\delta_\ell^*&=&-\frac{\pi e^2R_\odot}{m_ec^2k}\sum_i\frac{\alpha_i\beta_i}{\beta_i-1}B[{\textstyle\frac{1}{2}}\beta_i+{\textstyle\frac{1}{2}},{\textstyle\frac{1}{2}}]\Big(\frac{R_\odot}{b}\Big)^{\beta_i-1}.
\label{eq:a_b_del-r+-tot*}
\end{eqnarray}
Following \cite{Turyshev-Toth:2018-plasma} and using the phenomenological coefficients $\alpha_i, \beta_i$ listed after the model (\ref{eq:n_n-ss}), we estimate the angle of light deflection by the solar plasma, $\delta\theta_{\tt p}={d \delta^*_\ell}/{d \ell}$, as a function of the impact parameter and the wavelength:
{}
\begin{eqnarray}
\delta\theta_{\tt p}&=&\Big\{
6.62\times 10^{-13}\Big(\frac{R_\odot}{b}\Big)^{16}+
2.05\times 10^{-13}\Big(\frac{R_\odot}{b}\Big)^{6}\nonumber\\
&&\hskip 80pt +\,
2.42\times 10^{-16}\Big(\frac{R_\odot}{b}\Big)^2\Big\}\Big(\frac{\lambda}{1~\mu{\rm m}}\Big)^2.
\label{eq:ang*ip}
\end{eqnarray}
 For typical observing situations with reasonable Sun-Earth-probe separation angles \cite{Bertotti-Giampieri:1998,Verma-etal:2013}, expression (\ref{eq:ang*ip}) provides a good description. In fact, expressions (\ref{eq:a_b_del-r+-tot*})--(\ref{eq:ang*ip}) agree with those derived in \cite{Giampieri:1994kj,Bertotti-Giampieri:1998} and used in a recent test of general relativity  using radio links with the Cassini spacecraft  \cite{Bertotti-etal-Cassini:2003}.

\section{EM field in the interference region}
\label{sec:IF-region}

Now we are ready to present the components of the EM field in the interference region in the presence of plasma. In accord with (\ref{eq:Pi-g+p}), the total EM field in this region is given from (\ref{eq:alpha*1*})--(\ref{eq:A*B*}). To compute these factors, we replace the sums present in these expressions with integrals which are then evaluated by the method of stationery phase. As a result, to ${\cal O}(\theta^2,\delta\theta^2_{\tt p},r_g^2,\delta\theta_{\tt p}\sqrt{{2r_g}/{r}},(kr)^{-1})$, the total EM field in the interference region is given in the form
{}
\begin{align}
  \left( \begin{aligned}
{    D}_r& \\
{    B}_r& \\
  \end{aligned} \right) =& -iE_0
\sqrt{\frac{2r_g}{r} }\sqrt{2\pi kr_g}e^{i\sigma_0}J_1\Big(k\big(\sqrt{2r_gr}-r\delta\theta_{\tt p}\big) \theta\Big)\nonumber\\
&{}\times e^{i(kr+2\delta^\star_\ell -\omega t)}
\left( \begin{aligned}
\cos\phi \\
\sin\phi  \\
  \end{aligned} \right),
    \label{eq:DB-tot-rr}\\
  \left( \begin{aligned}
{    D}_\theta& \\
{    B}_\theta& \\
  \end{aligned} \right) =& E_0
 \sqrt{2\pi kr_g}e^{i\sigma_0}\Big(1-
 \frac{\delta\theta_{\tt p}}{\sqrt{2r_g/r}}\Big)J_0\Big(k\big(\sqrt{2r_gr}-r\delta\theta_{\tt p}\big) \theta\Big)\nonumber\\
 &{}\times e^{i(kr+2\delta^\star_\ell-\omega t)}\left( \begin{aligned}
\cos\phi \\
\sin\phi  \\
  \end{aligned} \right),
  \label{eq:DB-tot-th}\\
\left( \begin{aligned}
{    D}^{\tt }_\phi& \\
{    B}^{\tt }_\phi& \\
  \end{aligned} \right) =& E_0
  \sqrt{2\pi kr_g}e^{i\sigma_0}\Big(1-
  \frac{\delta\theta_{\tt p}}{\sqrt{2r_g/r}}\Big)J_0\Big(k\big(\sqrt{2r_gr}-r\delta\theta_{\tt p}\big) \theta\Big)\nonumber\\
  &{} \times e^{i(kr+2\delta^\star_\ell-\omega t)}\left( \begin{aligned}
-\sin\phi \\
\cos\phi  \\
  \end{aligned} \right).
  \label{eq:DB-tot-ph}
\end{align}

To study this field in the image plane, we need to transform (\ref{eq:DB-tot-rr})--(\ref{eq:DB-tot-ph})  to a cylindrical coordinate system \cite{Turyshev-Toth:2017} using the coordinate transformations $ \rho=R\sin\theta,$ $ z=R\cos\theta$. As a result, for a high-frequency EM wave (i.e., neglecting terms $\propto(kr)^{-1}$) and for $r\gg r_g$, we derive the field near the optical axis up to ${\cal O}(\rho^2/z^2)$ in the form
{}
\begin{align}
  \left( \begin{aligned}
{E}_z& \\
{H}_z& \\
  \end{aligned} \right) =&{\cal O}\Big(\frac{\rho}{z}\Big),
    \label{eq:DB-sol-z}\\
    \left( \begin{aligned}
{E}_\rho& \\
{H}_\rho& \\
  \end{aligned} \right) =&
 E_0
  \sqrt{2\pi kr_g}e^{i\sigma_0}\Big(1-
  \frac{\delta\theta_{\tt p}}{\sqrt{2r_g/r}}\Big)J_0\Big(k\sqrt{2r_gr}\Big(1-\frac{\delta\theta_{\tt p}}{\sqrt{2r_g/r}}\Big) \theta\Big)\nonumber\\
  &{}\times e^{i(kr+2\delta^\star_\ell-\omega t)}
 \left( \begin{aligned}
 \cos\phi& \\
 \sin\phi& \\
  \end{aligned} \right),
  \label{eq:DB-sol-rho}\\
    \left( \begin{aligned}
{E}_\phi& \\
{H}_\phi& \\
  \end{aligned} \right) =&
E_0 \sqrt{2\pi kr_g}e^{i\sigma_0}\Big(1-
\frac{\delta\theta_{\tt p}}{\sqrt{2r_g/r}}\Big)J_0\Big(k\sqrt{2r_gr}\Big(1-\frac{\delta\theta_{\tt p}}{\sqrt{2r_g/r}}\Big) \theta\Big)\nonumber\\
&{}\times e^{i(kr+2\delta^\star_\ell-\omega t)}
 \left( \begin{aligned}
 -\sin\phi& \\
 \cos\phi& \\
  \end{aligned} \right),
  \label{eq:DB-ph*}
\end{align}
where $r=\sqrt{z^2+\rho^2}=z(1+{\rho^2}/{2z^2})=z+{\cal O}(\rho^2/z)$) and $\theta=\rho/z+{\cal O}(\rho^2/z^2)$. These expressions are valid for forward scattering when $\theta\approx 0$, or when $\rho\leq r_g$.

Using the result (\ref{eq:DB-sol-z})--(\ref{eq:DB-ph*}), we may now compute the energy flux at the image region of the SGL. The relevant components of the time-averaged Poynting vector for the EM field in the image volume, as a result, may be given in the following form (see \cite{Turyshev-Toth:2017} for details):
{}
\begin{align}
{\bar S}_z=&\frac{c}{8\pi}E_0^2
\frac{4\pi^2}{1-e^{-4\pi^2 r_g/\lambda}}
\frac{r_g}{\lambda}\, \Big(1-\frac{\delta\theta_{\tt p}}{\sqrt{2r_g/z}}\Big)^2J^2_0\Big(2\pi\frac{\rho}{\lambda}\Big(\sqrt{\frac{2r_g}{z}}-\delta\theta_{\tt p}\Big)\Big),
\label{eq:S_z*6z}
\end{align}
with ${\bar S}_\rho= {\bar S}_\phi=0$ for any practical purposes. Also, we recognized that the following convenient expression is valid:
{}
\begin{eqnarray}
k\sqrt{2r_gr}\,\theta=2\pi\frac{\rho}{\lambda}\sqrt{\frac{2r_g}{z}}+{\cal O}(\rho^2/z).
\label{eq:J0}
\end{eqnarray}
Therefore, the non-vanishing component of the amplification vector $ {\vec \mu}$, defined as ${\vec \mu}={\vec {\bar S}}/|{\vec{\bar S}}_0|$ where $|{\bar {\vec S}}_0|=(c/8\pi)E_0^2$ is the time-averaged Poynting vector of the wave propagating in empty spacetime, takes the form
{}
\begin{align}
{\bar \mu}_z=&
\frac{4\pi^2}{1-e^{-4\pi^2 r_g/\lambda}}
\frac{r_g}{\lambda}\, \Big(1-\frac{\delta\theta_{\tt p}}{\sqrt{2r_g/z}}\Big)^2J^2_0\Big(2\pi\frac{\rho}{\lambda}\sqrt{\frac{2r_g}{z}} \Big(1-
\frac{\delta\theta_{\tt p}}{\sqrt{2r_g/z}}\Big)\Big).
\label{eq:S_z*6z-mu}
\end{align}

It is instructive to present  (\ref{eq:S_z*6z-mu}) in the following more informative form:
{}
\begin{eqnarray}
{\bar \mu}_z&=&
\frac{4\pi^2}{1-e^{-4\pi^2 r_g/\lambda}}
\frac{r_g}{\lambda}\, {\cal F}^2_{\tt pg}J^2_0\Big(2\pi\frac{\rho}{\lambda}
\sqrt{\frac{2r_g}{z}}
{\cal F}_{\tt pg}\Big),
\label{eq:S_z*6z-mu+}
\end{eqnarray}
where
\begin{eqnarray}
{\cal F}_{\tt pg}=\sqrt{1+\frac{\delta\theta^2_{\tt p}}{\delta\theta^2_{\tt g}}}-\frac{\delta\theta_{\tt p}}{\delta\theta_{\tt g}}\geq 0,
\end{eqnarray}
with $\delta\theta_{\tt g}=\sqrt{2r_g/z}=2r_g/b$ being the Einstein's deflection angle due to gravitational monopole. This result, to first order, is valid for any values of $\delta\theta_{\tt p}$ and $\delta\theta_{\tt g}$ and is very helpful to understand the impact of plasma on the optical properties of the SGL.  Although our analysis was conducted only to linear order in $\delta\theta_{\tt p}$, the presence of a quadratic term in the expression above is indicative of the overall behavior of the amplification factor ${\bar \mu}_z$  (\ref{eq:S_z*6z-mu+}).

As we can see from (\ref{eq:S_z*6z-mu+}), the plasma contribution to the optical properties of the SGL is governed by the factor ${\cal F}_{\tt pg}$, which, in the absence of plasma, is ${\cal F}_{\tt pg}=1$. Using $\delta\theta_{\tt g}=2r_g/b=8.49\times 10^{-6}\,(R_\odot/b)$ and $\delta\theta_{\tt p}$ from (\ref{eq:ang*ip}), we estimate the ratio of the two deflection angles as:
{}
\begin{align}
\frac{\delta\theta_{\tt p}}{\delta\theta_{\tt g}}=\Big\{&
7.80\times 10^{-8}\Big(\frac{R_\odot}{b}\Big)^{15}+
2.41\times 10^{-8}\Big(\frac{R_\odot}{b}\Big)^{5}\nonumber\\
&{}+
2.85\times 10^{-11}\Big(\frac{R_\odot}{b}\Big)\Big\}\Big(\frac{\lambda}{1~\mu{\rm m}}\Big)^2.
\label{eq:ang*ip+g}
\end{align}

Examining (\ref{eq:ang*ip+g}) as a function of the impact parameter, we see that for sungrazing rays passing by the Sun with impact parameter $b\simeq R_\odot$, this ratio reaches its largest value of $\delta\theta_{\tt p}/\delta\theta_{\tt g}=1.02\times 10^{-7} \,\big({\lambda}/{1~\mu{\rm m}}\big)^2$, which may be quite significant for microwave and longer wavelengths. For a wave with $\lambda\simeq 3$~mm passing that close to the Sun, the plasma contribution approaches that due to the gravitational bending, $\delta\theta_{\tt p}/\delta\theta_{\tt g}\sim 0.92$. As a result, the factor ${\cal F}_{\tt pg}$ from (\ref{eq:S_z*6z-mu+}) decreases to  ${\cal F}_{\tt pg}\sim 0.44$, which, as seen from (\ref{eq:S_z*6z-mu+}), leads to reducing the light amplification of the SGL to only ${\cal F}^2_{\tt pg}\sim 0.19$ compared to its value for the  plasma-free case and broadening the PSF by a  factor of ${\cal F}^{-1}_{\tt pg}\sim 2.28$, thus, reducing the angular resolution of the SGL in this case by the same amount. For the wavelength $\lambda\simeq 3$~cm, the ratio (\ref{eq:ang*ip+g}) increases to $\delta\theta_{\tt p}/\delta\theta_{\tt g}\sim 91.8$, which results in a light amplification factor of ${\cal F}^2_{\tt pg}\sim 2.97\times 10^{-5}$ compared to the plasma-free case with resolution degraded by the factor ${\cal F}^{-1}_{\tt pg}\sim184$.  Further increasing the wavelength to $\lambda\simeq 30$~cm leads to an obliteration of the optical properties of the SGL, where light amplification is reduced by a factor of $2.97\times 10^{-9}$ compared to the plasma-free case, with angular resolution degraded by the factor $1.84\times 10^{5}$.

At the same time, we can clearly see from (\ref{eq:ang*ip+g}) that for optical or IR bands, say for $\lambda\simeq 1~\mu$m or less, the ratio (\ref{eq:ang*ip+g}) is exceedingly small and may be neglected, which results in ${\cal F}_{\tt pg}= 1$ for waves in this part of the EM spectrum. This conclusion opens the way for using the SGL for imaging and spectroscopic applications of faint, distant targets.

\section{Discussion and Conclusions}
\label{sec:disc}

We studied the propagation of a monochromatic EM wave on the background of a spherically symmetric gravitational field produced by a gravitational mass monopole described in the first post-Newtonian approximation of the general theory of relativity taken and the solar corona represented by the free electron plasma distribution described by a generic, spherically symmetric power law model for the electron number density (\ref{eq:n_n-ss}). We used a generalized model for the solar plasma, which covers the entire solar system from the solar photosphere to the termination shock (i.e., valid for $0\leq r \leq R_\star$,   \cite{Turyshev-Toth:2018-plasma}). We considered the linear combination of gravity and plasma effects, neglecting interaction between the two. This approximation is valid in the solar system environment.

The static component of the solar plasma affects the optical properties of the SGL, especially for microwave or longer wavelengths. It leads to a  defocusing, which should not affect the size nor the position of the caustic line, except for the distance to the beginning of the focal line. Such plasma behavior does not induce aberrations, leaving the PSF of the SGL unchanged. Although temporal variability in the plasma may introduce additional aberrations, at optical wavelengths such effects are mostly negligible or may be accounted for with standard observational techniques \cite{Turyshev-Toth:2018a}. Short term temporal variability in the plasma may be accounted for by relying, for instance, on longer integration times. Alternatively, one may rely on the differential Doppler technique, introduced in \cite{Bertotti-Giampieri:1998} and applied in \cite{Bertotti-etal-Cassini:2003}, which would allow the plasma contribution to be greatly reduced, by more than three orders of magnitude.

In this approach, we relied on the spherical symmetry to capture the largest terms, representing the realistic field distributions in the solar system. An almost identical approach may be used to account for any nonsphericity that may be present either in the gravitational field or in the plasma distribution, or else would be introduced by imprecise spacecraft navigation and trajectory determination. Thus, the $1/r$ or $1/r^2$ terms may be included by applying the model that is already developed here. One would have to redefine the the $r_g$ and $\mu^2$ parameters in (\ref{eq:R-bar-k*2}). If quadrupole terms (i.e., terms in the potential that behave as $1/r^3$) are present, one can use a spherical coordinate system to solve the Maxwell equations. For higher order non-sphericity, given that for the solar system those terms are very small, one may develop a perturbation approach with respect to appropriately defined small parameters.

The approach presented here may be extended on a more general case of an extended Sun \cite{Roxburgh:2001,Park-etal:2017} and an arbitrary model of the solar plasma with a weak latitude dependence \cite{Muhleman-etal:1977,Muhleman-Anderson:1981}. We also emphasize that the SGL with its high light amplification and angular resolution properties could provide unique conditions to test relativistic gravitation and fundamental physics. This work is ongoing and results, when available, will be published elsewhere.

\begin{acknowledgements}
This work in part was performed at the Jet Propulsion Laboratory, California Institute of Technology, under a contract with the National Aeronautics and Space Administration.
\end{acknowledgements}

\bibliographystyle{spphys}       

\end{document}